\documentclass[aps]{revtex4}
\begin{document}
\title{Some intricacies of the momentum operator in quantum mechanics}
\author{Utpal Roy, Suranjana Ghosh}
\affiliation{Physical Research Laboratory, Ahmedabad 380009, India}
\author{Kaushik Bhattacharya}
\affiliation{Instituto de Ciencias, Universidad Nacional Aut\'{o}noma de 
M\'{e}xico, Circuito Exterior, C.U., A. Postal 70-543, C. Postal 04510, 
M\'{e}xico DF, M\'{e}xico}
\date{28:11:2007}
\begin{abstract}
In quantum mechanics textbooks the momentum operator is
defined in the Cartesian coordinates and rarely the form of the
momentum operator in spherical polar coordinates is
discussed. Consequently one always generalizes the Cartesian
prescription to other coordinates and falls in a trap. In this work we
introduce the difficulties one faces when the question of the momentum
operator in general curvilinear coordinates arises. We have tried to
elucidate the points related to the definition of the momentum
operator taking spherical polar coordinates as our specimen coordinate
system and proposed an elementary method in which we can ascertain the
form of the momentum operator in general coordinate systems.
\vskip .5cm
En los libros de mec\'{a}nica cu\'{a}ntica, el operador de
momento se define en coordenadas Cartesianas y raramente se discute la
forma de este operador en coordenadas polares. En consecuencia,
siempre se generaliza la prescripcion de este operador en coordenadas
Cartesianas al caso de otras coordenadas con lo cual se suele caer en
una trampa. En este trabajo, introducimos las dificultades que se
encuentran cuando surge la pregunta de como se escribe el operador de
momento en coordenadas curvilineas generales. Tratamos de dilucidar
los puntos relacionados con la definicion del operador de momento
tomado como ejemplo el caso de las coordenadas esfericas y proponemos
un m\'{e}todo elemental con el cual podemos establecer la forma del
operador de momento en sistemas coordenados generales.
\end{abstract}
\maketitle
\section{Introduction}
In classical mechanics the definition of momentum (both linear and
angular) in Cartesian coordinates is simple. Linear momentum is
defined as mass times the velocity and angular momentum is the
cross-product of the position vector with the linear momentum
vector, of a particle or a body in motion. In classical mechanics a
particle must have unique position and velocity and consequently the
definition of momentum is unambiguous.  When we are using
generalized coordinates then also the definition of the generalized
momenta are straight forward. We have to know the Lagrangian ${\cal
L}$ of the system written down in the generalized coordinates and
the momentum conjugate to the generalized coordinate $q_i$ is
simply:
\begin{eqnarray}
p_i \equiv \frac{\partial {\cal L}}{\partial \dot{q_i}}\,.
\label{genmom}
\end{eqnarray}
In quantum mechanics the position of a particle is not unique, one
has to revert to wave functions and then find out the probability
density of finding the particle in some portion of space. Naturally
the definition of momentum becomes a bit arbitrary. Elementary
textbooks of quantum mechanics \cite{element1, element2, element3}
invariably define the momentum operators in Cartesian coordinates
where the ambiguities are fortunately less. In Cartesian coordinates
we have three coordinates which have the same dimensions and linear
momentum operator is defined as:
\begin{eqnarray}
p_{i}=-i\hbar \frac{\partial}{\partial x_{i}}\,,
\label{momdef}
\end{eqnarray}
where $i=1,2,3.$ The angular momentum vectors are defined as:
\begin{eqnarray}
L=-i\hbar ({{\textbf r}} \times \nabla)\,. \label{angmdef}
\end{eqnarray}
All of the above definitions of the momentum operators seems to be
flawless in Cartesian coordinates. But soon it is realized that the
definitions above are not all fine if we have to generalize our
results to various coordinate systems. In this article we will
illustrate the problems of defining the momentum operators in general
curvilinear coordinates.

If we choose spherical polar coordinates then the difficulty we face
is that all the momentum components are not of the same status (as in
Cartesian coordinates), as one is a linear momentum the other two are
angular momenta.  More over in quantum mechanics we do not have a
relation corresponding to Eq.~(\ref{genmom}) to find out the momenta
in arbitrary circumstances.  The nice world of separate linear and
angular momenta vanish and we have to search how to define the momenta
in this new circumstances. Added on to these difficulties we also have
to care whether the defined momentum operators are actually
self-adjoint. In this article we will not speak about the
self-adjointness of the operators, for a better review on this topic
the readers can consult \cite{gjg}. In quantum mechanics when ever we
say about angular momenta in the back of our mind we conceive of the
generators of rotation which follow the Lie algebra.  But strictly
speaking these fact is not true. In spherical polar coordinates
$[p_\theta\,,\,p_\phi]=0$ although both of them are angular momentum
operators. The Lie algebra of the angular momenta follows only when we
are working in the Cartesian coordinate.

In the present article we try to formulate the important properties of
the momentum operator which can be generalized to non-Cartesian
coordinates and in this process we point out which properties cannot
be generalized. We try to treat both angular and linear momenta on the
same footing and try to find out the properties of these operators. In
the following section we start with a general discussion of the
momentum operator. Sec \ref{defmom} is dedicated to describe the
momentum operator in the Cartesian and the spherical polar coordinate
systems but its content is general and can be used to understand the
form of the momentum operators in other coordinates as well. We
conclude with brief discussion on the topics described in the article
in section \ref{cpc}.
\section{The momentum operator}
\label{gendisc}
Particularly in this section when we speak of momentum we will not
distinguish between linear or angular momenta. The coordinate system
in which the position and momentum operators are represented is
general with no bias for Cartesian system. The basic commutation
relations in quantum mechanics are:
\begin{eqnarray}
[q_i\,,\,p_j] = i\hbar \delta_{ij}\,,
\label{qp}
\end{eqnarray}
and
\begin{eqnarray}
[q_i\,,\,q_j]= [p_i\,,\,p_j] = 0
\label{qqpp}
\end{eqnarray}
where $q_j$ and $p_j$ are the generalized coordinate and momentum
operator and $\delta_{i\,j}=1$ when $i=j$ and zero for all other
cases, and $i,j=1,2,3$. From Eq.~(\ref{qp}) we can infer that the most
general form of the momentum operator in quantum mechanics, in
position representation, is:
\begin{eqnarray}
p_i = -i\hbar \left[\frac{1}{f(q)}\frac{\partial}{\partial q_i}
f(q) + h_{i}(q) + c_i \right]\,, 
\label{mgp}
\end{eqnarray}
where $f(q)$ and $h_i(q)$ are some arbitrary functions of coordinates
$q$ and $c_i$ is a constant all of which may be different for
different components of the momenta $p_i$. If we put this general form
of the momentum operator in Eq.~(\ref{qqpp}) we see that it restricts
$h(q)$ to be a constant and so the general form of the momentum
operator must be:
\begin{eqnarray}
p_i = -i\hbar \left[\frac{1}{f(q)}\frac{\partial}{\partial q_i}
f(q) + c_i \right]\,.
\label{gpo}
\end{eqnarray}
The form of $f(q)$ is arbitrary and has to be determined in
different circumstances.  In this article we show that one of the ways
in which the function $f(q)$ can be determined is imposing the
condition that the momentum operator must have a real expectation
value. The expectation value of the momentum operator is given as:
\begin{eqnarray}
\langle p_i \rangle \equiv \int \sqrt{-g(q)}\,
\psi^*(q) \,p_i \,\psi(q)\,\,
d^3 q\,,
\label{med}
\end{eqnarray}
where $g(q)$ is the determinant of the metric of the three dimensional
space.  The constant $c_i$ in the general form of the momentum
operator in Eq.~(\ref{gpo}) turns out to be zero. Before going into
the actual proof of the last statement it is to be noted that in any
arbitrary coordinate system some of the canonical conjugate momenta
will be linear and some will be angular. As the expectation values of
the canonical momentum operators must be real so for bound quantum
states the expectation value of the linear momentum operators must be
zero in any arbitrary coordinate system. Physically this means that
coordinates with linear dimensions are in general non-compact and
extend to infinity and for bound states if the expectation value of
the momenta conjugate to those non-compact coordinates are not zero
then it implies a momentum flow to infinity which contradicts the very
essence of a bound state. On the other hand in an arbitrary coordinate
system the expectation values of the angular momentum operators for
any bound/free quantum state can be zero or of the form $M \hbar$
(from purely dimensional grounds) where $M$ can be any integer
(positive or negative) including zero. To prove that $c_i=0$ we first
take those components of the momentum operator for which its
expectation value turns out to be zero in an arbitrary coordinate
system for a bound quantum state. Choosing those specific components
of the momentum we assume that for these cases $c_i \ne 0$ and it has
a unique value. The value of $c_i$ may depend on our coordinate choice
but in any one coordinate system it must be unique. Suppose in the
special coordinate system $q$ the normalized wave function of an
arbitrary bound state is given by $\psi(q)$. The from of the
definition of the general momentum operator in Eq.~(\ref{gpo}) then
implies:
\begin{eqnarray}
\langle p_i \rangle = 0 = -i\hbar F[\psi(q)] + c_i\,,
\label{psiexp}
\end{eqnarray}
where $F[\psi(q)]$ is a functional of $\psi(q)$ given by:
\begin{eqnarray}
F[\psi(q)]=\int \sqrt{-g(q)}\,
\psi^*(q)\,\left[\frac{1}{f(q)}\frac{\partial}{\partial
  q_i}\, \{f(q) \psi(q)\}\right]\,d^3 q\,,
\end{eqnarray}
which depends on the functional form of $\psi(q)$. If we choose
another bound state in the same coordinate system whose normalized
wave function is given by $\phi(q)$ then we can also write
\begin{eqnarray}
-i\hbar F[\phi(q)] + c_i = 0\,.
\label{phiexp}
\end{eqnarray}
Comparing Eq.~(\ref{psiexp}) and Eq.~(\ref{phiexp}) we see that the
value of $c_i$ does not remain unique. The only way out is that $c_i$
must be zero at least for those components of the momentum operator
which has vanishing expectation values in bound states. Next we take
those components of momentum operator whose expectation values are of
the form $M\hbar$. In these case also we can take two arbitrary but
different bound state wave-functions $\psi(q)$ and
$\phi(q)$. Evaluating $\langle p_i \rangle$ using these two different
wave-functions will yield two similar spectra of momenta as $M\hbar$
and $N\hbar$ where $M$ and $N$ belongs to the set of integers
including zero. In this case also we see that two quantum systems with
different wave-functions yielding similar momentum expectation
values. Consequently in this case also we must have $c_i=0$ other wise
it will not be unique.  Although we utilized some properties of bound
systems in quantum mechanics to show that $c_i=0$ in general but we
assume that the form of the canonical momentum operator is the same
for bound and free quantum states. Thus the most general form of the
canonical momentum operator for any quantum mechanical system in any
coordinate system must be of the form:
\begin{eqnarray}
p_i = -i\hbar \frac{1}{f(q)}\frac{\partial}{\partial q_i}f(q)\,.
\label{mgpo}
\end{eqnarray}
which still contains the arbitrary function $f(q)$ whose form is
coordinate system dependent.  From Eq.~(\ref{mgp}) we dropped the
function $h_i(q)$ because the momenta components must satisfy
$[p_i\,,\,p_j] = 0$. The last condition is not always true as in the
case of charged particles in presence of an external classical
electromagnetic field. For charged particles in presence of an
classical external electromagnetic field the momenta components do not
commute and in those cases the most general form of the momentum
operator will contain $h_i(q)$ as in Eq.~(\ref{mgp}) while $c_i$ will
be zero. In presence of an electromagnetic field the momenta
components of the electron are defined in the Cartesian coordinates as
$\pi_i \equiv -i\hbar \frac{\partial}{\partial x_{i}} + ie A_i({\bf
  x})$, where $e$ is the negative electronic charge and $A_i({\bf x})$
is the electromagnetic gauge field. In this case
$[\pi_i\,,\,\pi_j]=-ieF_{ij}$ where $F_{ij}$ is the electromagnetic
field-strength tensor. As seen in these cases $h_i(q)$ will be
proportional to the electromagnetic gauge field.

In this article we assume that the form of the momentum operator in
general orthogonal curvilinear coordinate systems is dictated by the
basic commutation relations as given in Eq.~(\ref{qp}) and
Eq.~(\ref{qqpp}) and the condition that the expectation value of the
momentum operator must be real in any coordinate system. To find out
the form of $f(q)$ we will follow an heuristic method. A more formal
approach to something similar to the topics discussed in this article
can be found in \cite{pod, dewitt}.  As the present article is purely
pedagogical in intent so we will try to follow the path as (probably)
taken by young graduate students who generalize the concepts of the
Cartesian coordinates to any other coordinate system and doing so
falls in a trap. Then after some thinking the student understands the
error in his/her thought process and mends his/her way to proceed
towards the correct result. Consequently, the ansatz which we will
follow to find out the form of $f(q)$ is the following. First we
will blindly assume, in any specific coordinate system, that
$f(q)=1$ as in the Cartesian coordinates and try to see whether the
momentum operator yields a real expectation value. If this choice of
$f(q)$ produces a real expectation value of the momentum operator
then the choice is perfect and we have the desired form of the
momentum operator. If on the other hand with our initial choice of
$f(q)=1$ we do not get a real expectation value of the momentum
operator then we will choose the appropriate value of it such that the
redefined momenta yields real expectation values. In the cases which
we will consider in this article the form of $f(q)$ will be evident
as soon as we demand that the momentum operators must have real
expectation values.

In this article we assume that in general the wave functions, we deal
with, are normalized to unity and the coordinate systems to be
orthogonal. More over the wave functions are assumed to vanish at the
boundaries or satisfy periodic boundary conditions.
\section{The momentum operator in various coordinate systems}
\label{defmom}
\subsection{Cartesian coordinates}
According to our ansatz here we initially take $f(x,y,z)=1$ which
implies that the $x$ component of the momentum operator is as given in
Eq.~(\ref{momdef}). If the normalized wave function solution of the
time-independent Schr\"{o}dinger equation of any quantum system is
given by $\psi({\bf x})$, where $\psi({\bf x})$ vanishes at the
boundaries of the region of interest, the expectation value of ${p}_x$
is:
\begin{eqnarray}
\int\psi^*({\bf x})\,{p}_x\,\psi({\bf x})\,d^3 x
&=& -i\hbar\int \int dy\,dz\,\left[\int_{-L}^L \psi^*({\bf x},t)\,
\frac{\partial\psi({\bf x})}{\partial x}\,dx \right]\nonumber\\
&=& -i\hbar \int \int dy\,dz\,\left[\left.\psi^*({\bf x})\psi({\bf x})
\right|^L_{-L} - \int^L_{-L}\psi({\bf x})
\frac{\partial\psi^*({\bf x})}{\partial x}\,dx \right]\,,
\label{momexp}
\end{eqnarray}
If the wave function vanishes at the boundaries $L$ and $-L$ or are
periodic or anti-periodic at those points then the first term on the
second line on the right-hand side of the above equation drops and we
have,
\begin{eqnarray}
\int\psi^*({\bf x})\,{p}_x\,\psi({\bf x})\,d^3 x
&=& -i\hbar \int \int dy\,dz\,\left[\int^L_{-L} \psi^*({\bf x})\,
\frac{\partial\psi({\bf x})}{\partial x}\,dx \right]\nonumber\\
&=& i\hbar \int \int dy\,dz\,\left[\int^L_{-L}\psi({\bf x})
\frac{\partial\psi^*({\bf x})}
{\partial x}\,dx\right]\nonumber\\
&=&\int \psi({\bf x})\,p_x^*\,\psi^*({\bf x})\,d^3 x\,.
\label{hermomexp}
\end{eqnarray}
This shows that the expectation value of ${p}_x$ is real and so in
Cartesian coordinates we have $f(x,y,z)=1$. In a similar way it can
be shown that with the same choice of $f(x,y,z)$ 
the expectation values of ${p}_y$ and ${p}_z$ are also real. The above
analysis also can be done when the wave functions are separable.
Consequently in Cartesian coordinates the momentum operators are given
as in Eq.~(\ref{momdef}).
\subsection{Spherical polar coordinates}
\label{intri}
\subsubsection{The radial momentum operator}
If we start with $f(r,\theta,\phi)=1$ then the radial
momentum operator looks like:
\begin{eqnarray}
p'_r= -i\hbar \frac{\partial}{\partial r}\,.
\label{wrdef}
\end{eqnarray}
If the solution of the time-independent Schr\"{o}dinger equation for
any particular potential be $\psi({r,\theta,\phi})\equiv \psi({\bf r})$ then
the expectation value of $p'_r$ is,
\begin{eqnarray}
\langle p'_r \rangle = -i\hbar \int d\Omega \left[ 
\int^{\infty}_0 r^2 \,\psi^*({\bf r})\,\,\frac{\partial \,\psi({\bf r})}
{\partial r}\,dr 
\right]\,,
\end{eqnarray}
where $d\Omega=\sin\theta \,d\theta \,d\phi$ and consequently,
\begin{eqnarray}
\langle {p'}_r \rangle &=& -i\hbar \int d\Omega \left[
\int^{\infty}_0 r^2 \psi^*({\bf r})\frac{\partial \psi({\bf r})}{\partial r}
\,dr \right]\nonumber\\
&=&-i\hbar \int d\Omega \left[ \left.r^2 \psi^*({\bf r}) \psi({\bf r})
\right|^\infty_0  -
\int^\infty_0 \left(2r \psi^*({\bf r}) + r^2 \frac{\partial \psi^*({\bf r})}
{\partial r}\right)\psi({\bf r})\,dr
\right]\,.
\label{rherm}
\end{eqnarray}
If $\psi({\bf r})$ vanishes as $r \to \infty$ 
then the above equation reduces to,
\begin{eqnarray}
\langle {p'}_r \rangle &=& i\hbar \int d\Omega \left[\int^\infty_0 r^2 
\psi({\bf r}) \frac{\partial \psi^*({\bf r})}{\partial r}\,dr \right] + 
2 i\hbar \int d\Omega \int^\infty_0 r |\psi({\bf r})|^2 \,dr\nonumber\\
&=& \langle {p'}_r \rangle^* + 2 i\hbar \int d\Omega \int^\infty_0 r 
|\psi ({\bf r})|^2 \,dr\,.
\label{nonherm}
\end{eqnarray}
The above equation implies that $\langle {p'}_r\rangle$ is not real in
spherical polar coordinates. On the other hand if we write
Eq.~(\ref{nonherm}) as:
\begin{eqnarray}
\langle {p'}_r \rangle - i\hbar \int d\Omega \int^\infty_0 r 
|\psi({\bf r}))|^2 \,dr
= \langle {p'}_r \rangle^* +  i\hbar \int d\Omega \int^\infty_0 r 
|\psi({\bf r}))|^2 \,dr\,,
\label{nonhermi}
\end{eqnarray}
then the left hand side of the above equation can be written as:
\begin{eqnarray}
\langle {p'}_r \rangle - i\hbar \int d\Omega \int^\infty_0 r 
|\psi({\bf r})|^2 \,dr
&=& -i \hbar \int d\Omega \int^{\infty}_0 \left[r^2 \psi^*({\bf r})
\frac{\partial \psi({\bf r})}{\partial  r}
+ r |\psi({\bf r})|^2\right]\,dr\nonumber\\
&=& -i \hbar \int d\Omega \int^{\infty}_0 r^2 \psi^*({\bf r})
\left[ \frac{\partial}{\partial r} + \frac{1}{r}
\right]\psi({\bf r})\,dr\,.
\end{eqnarray}
A similar manipulation on the right side of Eq.~(\ref{nonhermi}) can be done
and it yields:
\begin{eqnarray}
\langle {p'}_r \rangle^* + i\hbar \int d\Omega \int^\infty_0 r 
|\psi ({\bf r})|^2 \,dr
&=& i \hbar \int d\Omega \int^{\infty}_0 \left[r^2 \psi({\bf r})
\frac{\partial \psi^* ({\bf r})}{\partial r}
+ r |\psi({\bf r})|^2\right]\,dr\nonumber\\
&=& i \hbar \int d\Omega \int^{\infty}_0 r^2 \psi({\bf r})\left[ 
\frac{\partial}{\partial r} + \frac{1}{r}
\right]\psi^*({\bf r})\,dr\,.
\end{eqnarray}
Now if  we redefine the radial momentum operator as:
\begin{eqnarray}
{p}_r \equiv {p'}_r -\frac{i\hbar}{r} =-i\hbar\left
(\frac{\partial}{\partial r} +
\frac{1}{r}\right)= -i\hbar \frac{1}{r}\frac{\partial}{\partial r}\,r\,,
\label{remom}
\end{eqnarray}
then from Eq.(\ref{nonhermi}) we observe that the expectation value
of ${p}_r$ must have real values. This fact was derived in a
different way by Dirac \cite{dirac, sieg}.  Now this form of ${p}_r$
we can identify as $f(r,\theta,\phi)=r$ in Eq.\ (\ref{gpo}).

We end our discussion on the radial component of momentum operator
with an interesting example. Here we will calculate the expectation
value of the radial component of the force acting on an electron in the
Hydrogen atom. The main purpose of this exercise is to generalize
Heisenberg's equation of motion in spherical polar coordinates which
is not found in the elementary quantum mechanics text books. Although
in the Heisenberg picture we deal with operators but to specify the
relevant quantum numbers we first write down the wave-function of the
Hydrogen atom explicitly. In the present case the wave-function is
separable and it is given as:
\begin{eqnarray}
\psi_{n \, L \, M}(r,\theta,\phi)
&= & N_r \, R_{n \, L}(r) \, Y_{L \, M}(\theta,\phi)\,,\nonumber\\
& = & N_r \, e^{-r/n a_0}
\left[\frac{2r}{n a_0} \right]^{L} {\cal L}_{n - L -1}^{2 L + 1}
\left(\frac{2r}{n a_0} \right) Y_{L \, M}(\theta,\phi)\,,
\label{hatsol}
\end{eqnarray}
where $a_0=\frac{\hbar^2}{m e^2}$ is the Bohr radius and $m$ is the
reduced mass of the system comprising of the proton and the electron.
$n$ is the principal quantum number which is a positive integer,
${\cal L}_{n - L -1}^{2 L + 1}(x)$ are the associated Laguerre
polynomials, $Y_{L \, M}(\theta,\phi)$ are the spherical-harmonics,
and $N_r$ is the normalization arising from the radial part of the
eigenfunction.  The domain of $L$ is made up of positive integers
including zero and $M$ are such that for each $L$, $-L \le M \le
L$.  The radial normalization constant is given by:
\begin{equation}
N_r = \left[ \left(\frac{2}{n a_0} \right)^3 \frac{(n - L -
1)!}{(n + L)! 2n} \right]^{1/2}.
\end{equation}
The spherical-harmonics are given by,
\begin{equation}
Y_{L \, M}(\theta,\phi) = (-1)^M\,\left[\frac{2 \, L + 1}{4 \pi}
\frac{(L - M)!}{(L + M)!} \right]^{1/2} P^L_M(\cos \theta)
e^{i M \phi}\,,
\end{equation}
where $P_L^M(\cos \theta)$ are the associated Legendre functions. 

From elementary quantum mechanics textbooks we know that in Cartesian
coordinates the time evolution of the momentum operator in one
dimension is given as:
\begin{eqnarray}
\frac{d {p}_x}{dt}=\frac{1}{i\hbar}[{p}_x, {H}]
= -\frac{d}{d x} V({x})\,,
\label{nlaw}
\end{eqnarray}
which is the operator version of Newton's second law. Now if we take
the expectation values of both sides of Eq.~(\ref{nlaw}) in any basis
we get:
\begin{eqnarray}
\frac{d \langle {p}_x \rangle }{dt}
= -\left\langle \frac{d}{d x} V({x}) \right\rangle\,,
\label{eht}
\end{eqnarray}
which is called the Ehrenfest theorem. In the case of the Hydrogen
atom $V({\bf r})=-\frac{e^2}{r}$ and if we follow the Cartesian
prescription we will write,
\begin{eqnarray}
\frac{d {p}_r}{d t}= - \frac{d}{d r} V({r})
= - \frac{e^2}{r^2}\,,
\label{wrng1}
\end{eqnarray}
which implies that the only force acting on the electron is the
centripetal force supplied by the Coulomb field. But the above
equation has some difficulties. First from our knowledge of particle
mechanics in central force fields we know that there must be some
centrifugal reaction also which is absent in
Eq.~(\ref{wrng1}). Secondly as the Hydrogen atom is a bound system the
expectation value of its radial component of the momentum vanish and
consequently the time rate of change of the radial momentum
expectation value must also vanish. But if we take the expectation
value of the right hand side of Eq.~(\ref{wrng1}) it does not
vanish. So Eq.~(\ref{wrng1}) is a wrong equation and we cannot blindly
use the Cartesian prescription in such a case.

The correct way to proceed in the present circumstance is to write
Heisenberg's equation of motion in spherical polar coordinates using
the Eq.~(\ref{remom}) as the radial momentum operator.  In this way we
will get all the results right. The Hamiltonian of the Hydrogen atom
is:
\begin{eqnarray}
{H}= -\frac{\hbar^2}{2 m} \frac{1}{r}\frac{\partial^2}{\partial r^2}\,r
+ \frac{1}{2 m r^2} {\bf L}^2 - \frac{e^2}{r}\,,
\label{hham}
\end{eqnarray}
where,
\begin{eqnarray}
{\bf L}^2 = -\hbar^2\left(\frac{1}{\sin\theta} \frac{\partial}{\partial
\theta}\sin\theta\frac{\partial}{\partial \theta} + \frac{1}{\sin^2\theta}
\frac{\partial^2}{\partial \phi^2}\right)\,,
\end{eqnarray}
whose eigenvalues are of the form $\hbar^2 L(L+1)$ in the basis $Y_{L
\, M}(\theta, \phi)$. In the expression of the Hamiltonian $m$ is the
reduced mass of the system comprising of the proton and electron. Next
we try to apply Heisenberg's equation to the radial momentum
operator. Noting that the first term of the Hamiltonian is nothing but
${p}^2_r$ the Heisenberg equation is:
\begin{eqnarray}
\frac{d {p}_r}{d t} &=&
- \frac{{\bf L}^2}{2 m}\left[\frac{1}{r}\frac{\partial}{\partial r}\,
r\,,\,\frac{1}{r^2}\right] + e^2\left[\frac{1}{r}\frac{\partial}
{\partial r}\,r\
\,,\,\frac{1}{r}\right]\,,\nonumber\\
&=& \frac{{\bf L}^2}{m r^3} - \frac{e^2}{r^2}\,.
\label{hheis}
\end{eqnarray}
The above equation is the operator form of Newton's second law in
spherical polar coordinates. The first term in the right hand side of
the above equation gives the centrifugal reaction term in a central
force field. Evaluating the expectation value of both the sides of the
above equation using the wave-functions given in Eq.~(\ref{hatsol}) we
get \cite{element1},
\begin{eqnarray}
\left\langle \frac{1}{r^2} \right\rangle &=& \frac{1}{n^3\, a^2_0 (L +
\frac12)}\,,\\
\left\langle \frac{1}{r^3} \right\rangle &=& \frac{1}{a^3_0 \, n^3 L (L +
\frac12) (L+1)}\,.
\end{eqnarray}
Using the above expectation values in Eq.~(\ref{hheis}) and noting
that $\langle {\bf L}^2 \rangle = \hbar^2 L (L + 1)$ we see that the
time derivative of the expectation value of the radial momentum
operator of the Hydrogen atom vanishes.  The interesting property to
note is that although the Heisenberg equation of motion for ${p}_r$
shows that a force is acting on the system due to which ${p}_r$ is
changing but as soon we go to the level of expectation values the
force equation collapses to give a trivial identity. The cause of this
is the reality of the radial momentum operator which is bound to have
a real value. 
\subsubsection{The angular momentum operator canonically conjugate to $\phi$ 
and $\theta$}
\label{phi}
The canonically conjugate momenta corresponding to the angular
variables must be angular momentum operators. Let ${p}_\phi$ be
the angular momentum operator canonically conjugate to $\phi$. In this
case if we set $f(r,\theta,\phi)=1$ the form of the momentum
operator is:
\begin{eqnarray}
p_\phi=-i\hbar\frac{\partial}{\partial \phi}\,,
\label{lphi}
\end{eqnarray}
which can be shown to posses real expectation values by following a
similar proof as is done in Eq.~(\ref{momexp}) and
Eq.~(\ref{hermomexp}), if we assume $\Phi(0)=\Phi(2\pi)$. The periodic
boundary condition forces the solution of the time-independent
Schr\"{o}dinger equation to be of the form:
\begin{eqnarray}
\psi({r,\theta,\phi}) \propto \frac{1}{\sqrt{2\pi}}\,e^{iM \phi}\,,
\end{eqnarray}
where $M=0,\pm 1,\pm 2, \cdot \cdot$. The solution cannot have any
other form of $M$ dependence as in that case the expectation value of
$L_\phi$ will not be of the form $M\hbar$. So for the case of $\phi$
our initial choice of $f(r,\theta,\phi)=1$ turns out to be correct.

Next we compute the expectation value of ${p}_\theta$ with
$f(r,\theta,\phi)=1$. The expectation value of $\langle {p}_\theta
\rangle$ is as follows:
\begin{eqnarray}
\langle {p}_\theta \rangle &=& -i\hbar \int \int r^2 dr d\phi
\left[\int^{\pi}_0 \psi^*({\bf r})
\frac{\partial \psi({\bf r})}{\partial \theta}\,\sin\theta d\theta
\right]\nonumber\\
&=& -i\hbar \int \int r^2 dr d\phi \left[\left.\sin\theta \,\psi^*({\bf r})
\psi({\bf r})\right|^\pi_0 -  \int^{\pi}_0 \left(
\cos\theta \,\psi^*({\bf r}) + \sin\theta
\frac{\partial\psi^*({\bf r})}{\partial\theta}\right)\psi({\bf r})\,
d\theta\right]
\nonumber\\
&=& \int \int r^2 dr d\phi \left[i\hbar\int^{\pi}_0 \sin\theta\,\psi({\bf r})
\frac{\partial\psi^*({\bf r})}{\partial\theta}\,d\theta \right]
+ i\hbar \int \int r^2 dr d\phi \int^{\pi}_0 \cos\theta\,|\psi({\bf r})|^2\,
d\theta\nonumber\\
&=& \langle {L}_\theta \rangle^* + i\hbar \int \int r^2 dr d\phi 
\int^{\pi}_0 \cos\theta\,|\psi({\bf r})|^2\,d\theta\,.
\end{eqnarray}
The above equation shows that $\langle {p}_\theta\rangle$ is not real
and so our choice of $f(r,\theta,\phi)$ is not correct.  Following
similar steps as done for the radial momentum operator we can redefine
the angular momentum operator conjugate to $\theta$ as \cite{essen}:
\begin{eqnarray}
{p}_\theta \equiv -i\hbar\left(\frac{\partial}{\partial \theta}
+ \frac12 \cot\theta \right)\,.
\end{eqnarray}
From the form of ${p}_\theta$ we find that in this case
$f(r,\theta,\phi)=\sqrt{sin\theta}$ in Eq.\ (\ref{gpo}).
\subsubsection{The signature of the unequal domain of the angular variables 
in the spherical polar coordinates}
\label{ud}
It is known that both $\theta$ and $\phi$ are compact variables,
i.e. they have a finite extent. But there is a difference between
them. In spherical polar coordinates the range of $\phi$ and $\theta$
are not the same, $0 \leq \phi < 2\pi$ and $0 \leq \theta \leq
\pi$. This difference has an interesting outcome. As $\phi$ runs over
the whole angular range so the wave-function has to be periodic in
nature while due to the range of $\theta$ the wave-function need not
be periodic. Consequently there can be a net angular momentum along
the $\phi$ direction while there cannot be any net angular momentum
along $\theta$ direction. What ever the quantum system it may be we
will always have the expectation value of ${p}_\theta$ to be zero
although the wave-functions may not be periodic in $\theta$ or it may
not vanish at $\theta=0$ and $\theta=\pi$. In this article we will
show the validity of this observation for the cases where the
wave-functions are separable but the result holds for non-separable
wave-functions also.

As the time-independent Schr\"{o}dinger equation for a central
potential yields the wave-function corresponding to $\theta$ to be
$\Theta(\theta)$ which is given by:
\begin{eqnarray}
\Theta(\theta)=N_\theta\,P^L_M(\cos\theta)\,,
\label{thetaeqn}
\end{eqnarray}
where $N_\theta$ is a normalization constant depending on $L$, $M$ and
$P^L_M(\cos\theta)$ is the associated Legendre function, which is
real. In the above equation $L$ and $M$ are integers where $L=0,1,2,3,
\cdot \cdot$ and $M=0, \pm 1, \pm 2, \pm 3, \cdot \cdot$.  A
requirement of the solution of the Schr\"{o}dinger equation for a
central potential is that $-L \leq M \leq L$. Now we can calculate the
expectation value of ${p}_\theta$ using the above wave-function
and it is:
\begin{eqnarray}
\langle {p}_\theta \rangle &=& -i\hbar N^2_\theta\int^\pi_0
P^L_M(\cos\theta) \left(\frac{d P^L_M(\cos\theta)}{d\theta} +
\frac12 \cot\theta P^L_M(\cos\theta)\right)
\sin\theta d\theta\,\nonumber\\
&=& -i\hbar N^2_\theta\left[\int^\pi_0
P^L_M(\cos\theta)\frac{d P^L_M(\cos\theta)}{d\theta}
\sin\theta d\theta
+ \frac12 \int^\pi_0 P^L_M(\cos\theta) P^L_M(\cos\theta) \cos\theta \,d\theta
\right]\,.\nonumber\\
\end{eqnarray}
To evaluate the integrals on the right hand side of the
above equation we can take $x=\cos\theta$ and then the expectation
value becomes:
\begin{eqnarray}
\langle {p}_\theta \rangle &=& -i\hbar N^2_{\theta}\left[
\int^{-1}_1 P^L_M(x) \frac{d P_L^M (x)}{dx}
(1-x^2)^{\frac12}\,dx\right.\nonumber\\
&-&\left.\frac{1}{2}\int^{-1}_{1} P^{L}_{M(x)}
P^{L}_{M(x)}\frac{x}{\sqrt{1-x^{2}}} dx \right]\,. \label{postel}
\end{eqnarray}
The second term in the right hand side of the above equation
vanishes as the integrand is an odd function in the integration
range. For the first integral we use the following recurrence
relation \cite{grad}:
\begin{equation}
(x^2 - 1) \frac{d P^L_M (x)}{dx} =  M x P^L_M (x)
- (L+ M) P^L_{M-1}(x)\,,
\label{asslegen}
\end{equation}
the last integral can be written as,
\begin{eqnarray}
\langle {p}_\theta \rangle &=& i\hbar N^2_\theta\left[M \int^{-1}_1
x (1 - x^2)^{-\frac12} P^L_M(x)  P^L_M (x)\,dx \right.\nonumber\\
&-&\left. (L+M) \int^{-1}_1
(1 - x^2)^{-\frac12} P^L_M (x) P^L_{M-1}(x)\,dx \right]\,.
\label{le}
\end{eqnarray}
As,
\begin{eqnarray}
P^L_M(x)= (-1)^{L+M}P^L_M(-x)\,,
\label{ppar}
\end{eqnarray}
we can see immediately that both the integrands in the right hand
side of Eq.~(\ref{le}) are odd and consequently $\langle
{p}_\theta \rangle = 0$ as expected. A similar analysis gives
$\langle {p}_\phi \rangle = M\hbar$. As the motion along $\phi$ is
closed so there can be a net flow of angular momentum along that
direction but because the motion along $\theta$ is not so, a net
momentum along $\theta$ direction will not conserve probability and
consequently for probability conservation we must have expectation
value of angular momentum along such a direction to be zero.
\section{Conclusion}
\label{cpc}
Before we conclude we state that the form of the momentum operators in
the other widely used curvilinear coordinates as the cylindrical polar
coordinates or the plane polar coordinates can be found out similarly
as done for the spherical polar coordinates.  The form of the momentum
operator in most of the commonly used coordinates can be deduced from
the ansatz which we presented in section \ref{gendisc}. Although the
treatment presented in this article is not a general technique which
can be applied in all circumstance as in that case one has to proof
that this process of fixing the form of the momentum is unique and can
be applied for all coordinate systems, whatsoever pathological it may
be. 

The points discussed above are rarely dealt in elementary quantum
mechanics textbooks. Most often the linear momentum operator is
defined in Cartesian coordinates and it is intuitively attached to the
generator of translations. The difficulty of such an approach is that
it becomes very difficult to generalize such an approach to other
curvilinear coordinates where the concept of translation is
non-trivial. More over in arbitrary coordinate systems the concept of
the uncertainty of the position and momentum operators become a
difficult and non-trivial concept. As curvilinear coordinates contain
compact dimensions it may happen that the uncertainties in those
directions exceed the domain of the coordinate for a sharply defined
conjugate momentum and so the conventional understanding of the
uncertainty relation in general breaks down.

In the present work we have emphasized on the reality of the momentum
expectation value and using the reality of the expectation value as a
bench mark we did find out the form of the momentum. Linear and
angular momenta were not dealt differently.  This process of deduction
is interesting as in most of the cases the actual expectation value of
the momentum is zero. It may seem that the deductions were incorrect
as we manipulated zeros. In this regard it must be understood that the
actual expectation values of the momenta in bound state turns out to
be zero in many cases as because the expectation values requires to be
real. As in bound state problems we have real wave-functions so the
expectation value of the momenta operators can be real only when it
is zero. Consequently the reality of the expectation value is a
concept which is more important than the fact that in many cases the
expectation value turns out to be zero. In the present work the same
prescriptions which yield the forms of the linear momenta also gives
us the forms of the angular momenta. It was shown that irrespective of
the nature of the potential the expectation value of the angular
momentum conjugate to $\theta$ in spherical polar coordinates is
zero. This is more a geometric fact rather than a physical effect.
\begin{center}
{\bf Acknowledgements}
\end{center}
The authors thank Professors D. P. Dewangan, S. Rindani, J. Banerji,
P. K. Panigrahi and Ms. Suratna Das for stimulating discussions and
constant encouragements.

\end{document}